# Crystallization of β-phase Poly (vinylidene fluoride) films using dimethyl sulfoxide (DMSO) solvent and at suitable annealing condition

S. Satapathy[1], P. K. Gupta[1], Santosh Pawar[2] and K. B. R. Varma[3]

[1] Laser Materials Development & Devices Division, Raja Ramanna centre for Advanced Technology, Indore 452013, India

[2] Department of Applied Physics, SGSITS, Indore 452003, India

[3] Materials Research Centre, Indian Institute of Science, Bangalore 560012, India

E- mail: srinu73@cat.ernet.in

**Abstract:**

In literature it has been reported that the γ-phase PVDF film is formed from dimethyl sulfoxide (DMSO) solvents regardless of preparation temperature. In this report the crystallization of both α and γ phase from DMSO solvent by varying preparation temperature has been described. This paper also describes the conversion of γ-phase to β- phase and α-phase at different annealing condition. When thin films were annealed at $90^0 C$ for 5 hours, then maximum β-phase content (greater than 95%) is present in PVDF thin film. The PVDF films completely converted to α-phase, when they were annealed at $160^0 C$ for 5 hours. From (X-ray diffraction) XRD, Fourier Transform Infrared Spectrum (FTIR), (Differential thermal analysis) DTA and Raman studies, it is confirmed that the PVDF thin films, cast from DMSO solution and annealed at $90^0 C$ for 5 hours, have maximum percentage of β-phase(greater than 95%).

# 1. Introdiction

PVDF, a semi-crystalline polymer exists in four different forms. The β-phase is the desirable phase due to its ferroelectric characteristic. Phase I (β phase) has a planer TTTT(all trans) zigzag chain conformation which has space group Cm2m (orthorhombic, a= 8.58$A^0$, b= 4.91 $A^0$, c= 2.56 $A^0$) [1, 2]. The α-form, phase II (monoclinic, $P2_1/c$; a = 4.96 $A^0$, b= 9.64 $A^0$, c = 4.62 $A^0$, β = $90^0$) has a chain conformation which is approximately TGTG (trans- gauche-trans-gauche) [1, 2]. Phase III (γ-phase) has a chain conformation which is approximately TTTGTTTG$^/$, with the space group being C2cm [3]. Under application of high electric field to phase II (α-form), may cause $180^0$ rotation of alternate chains leading to a polar, monoclinic form IV crystal (δ-phase) [4-6]. Different phases exist in PVDF depending on synthesis conditions like solvent, melt temperature, method of casting, stretching of thin films and annealing conditions. M. G. Buonomenna et al. show the effect of casting, solvent and coagulation conditions on the crystalline structure of PVDF [7]

Dense β-phase is dominant in PVDF thin films which have been obtained by spin coating PVDF/DMF solution with addition of $Mg(NO_3)_2.6H_2O$ and dried at $100^0C$ [8]. Only below a certain temperature the deformation of the α-phase results in transformation to the β-phase (polar) form. B. Mohammadi et al. has shown the effect of stretch rate on conversion of α-phase to β-phase [9, 10]. Maximum β-phase content was obtained at $90^0C$ during stretching [11]. The abrupt increase in orientation of β crystallites and maximum content of the β form occur closely above the temperature of $α_c$ relaxation temperature (between 70-$87^0C$) [11].

The γ-phase has been produced by either casting from DMSO or dimethylformamide solution regardless of preparation temperature [12-16]. The γ-phase can also be produced by crystallizing from melt under high pressures [17]. The γ- phase can also be produced as a result of an atmospheric pressure annealing induced transformation from the α-phase [18]. The percentage of γ- phase has been increased when α-phase thin film annealed at $158^0C$ for 95hours [19]. When β-phase thin film is annealed at 181$^0$C at heating rate 0.31K/min, then large percentage of β-crystallite is transformed to γ-phase [19]. There is no clear report on the transformation of γ-phase to β phase in PVDF.

In this paper we report the crystallization of α and γ-phase PVDF from DMSO solvent at preparation temperature 50$^0$C and 90$^0$C respectively. We choose preparation temperatures above α-phase relaxation temperature (90$^0$C) and below α-phase relaxation temperature (50$^0$C) because above and below α-phase relaxation temperature different phases exist in PVDF films.

The conversion of γ- phase to β- phase (which is still unclear from literature studies) and γ- phase to α- phase at different annealing condition has also been reported. So in this manuscript it has been shown that the same solvent can yield the α, β and γ phases. Strong correlations have been observed among the characterization results. The variation of thin film properties due to annealing has been described comprehensively.

**2. Experiment**

PVDF thin films were prepared from DMSO solvent at two different temperatures (90$^0$C and 50$^0$C). The granular PVDF (Sigma Aldrich) was dissolved in the solvent DMSO (concentration is 15wt %). In first case the temperature of the solution was increased to 90$^0$C for complete dissolution of the polymer. The solution was cast uniformly on the glass substrate by means of hand casting at room temperature. The thin films were kept at room temperature for one hour and then dried at 60$^0$C. Free standing thin films were obtained after immersing in water.

In the second case, the PVDF (15 wt %) dissolved completely at 50$^0$C and cast at room temperature on glass plate. The thin films were kept it at room temperature for one hour and then dried at 60$^0$C.

The thin films (which were cast at room temperature from solution at 90$^0$C) were annealed at 70, 90, 110, 130 and 160$^0$C for 5 hours. Phases of thin films were analyzed by X-ray diffractometer, using Cu Kα radiation, with the generator working at 40kV and 40mA. FTIR spectra of the films were recorded by a Perkin Elmer Spectrometer over a range of 400cm$^{-1}$ to 1750cm$^{-1}$. The Raman measurements on PVDF films were carried out using 632.8nm excitation source using LABRAM-HR spectrometer equipped with a peltier cooled CCD detector. The melting behavior of PVDF films (annealed at different temperatures) were investigated by DTA (Setaram TGDTA model no.92) at a heating rate of 5$^0$C/min. The melting point, crystallization and phase transition of α-, β- and γ-phases of PVDF were observed from DTA thermo grams during heating cycle of DTA. Polarization (Hysteresis loop measurement) measurements of PVDF thin films were performed using RT66A Radiant ferroelectric loop tracer.

## 3. Results and Discussion

### I. Crystallization of α and γ-phase PVDF thin films from DMSO solvent at different solution temperatures.

The X-ray diffraction pattern of PVDF thin films, prepared from solutions at 90$^0$C and 50$^0$C (as described in experiment), are shown in figure1. From figure1 (a) it is observed that the maximum intensity γ-phase peak at 2θ = 20.3(101) [20] is mainly present in XRD pattern. So high percentage of γ-phase is present in PVDF thin film, those have been cast at room temperature from solution at 90$^0$C. But small α- phase peaks at 17.6 (100), 18.4(020), 19.9(110) and 26.6 (021)[6] peaks are also present in XRD pattern which indicates the existence of small percentage of α- phase along with γ-phase in thin film of PVDF cast at room temperature from solution at 90$^0$C.

In figure 1(b), the maximum intensity peak is observed at 19.9 (2θ values) along with peaks at 17.7 and 18.4, which are belongs to α-phase i. e. only α-phase exist in PVDF thin films, those have been cast at room temperature from solution at 50$^0$C.

**Why both α and γ phases of PVDF obtained at different preparation temperatures from solution of PVDF and DMSO?**

The existence of α, β and γ phases in the PVDF depends on the mobility of conformers which mainly affected by the thermal energy. In the β phase (polar phase) of PVDF the long trans zigzag segments are connected to each other with skew bond or equivalent gauche-trans sequences. The ferroelectric phase may be assumed as a kind of super lattice consisting of domains of long trans chains linked together with the boundaries of the disordered trans-gauch bonds. The α phase is a kind of phase in which gauch type chain rotates around the chain axis; therefore the total dipole moment becomes zero or a nonpolar crystal. The granular PVDF (Sigma Aldrich) when added to DMSO solvent at $50^0$C, the thermal energy is not sufficient to rotates the $CF_2$ group in turn there is no trans- gauch conformational change. So after casting the thin films remain in α-phase.

When solution prepared at $90^0$C using DMSO solvent, the thermal energy is enough to rotate $CF_2$ dipoles resulting a cooperative motion of the neighbouring $CF_2$ groups through the large scale trans-gauch conformational chage. So conformers are aligned either in TTTGTTTG' or TT form. So the solutions prepared using DMSO solvents at $90^0$C and immediately cast at room temperature produce polar PVDF films have large percentage of γ phase. So solvent does not matter to produce different phases of PVDF. Same solvent can yield α, β and γ phase.

**II. Formation of β- phase from γ-phase PVDF**

The γ-phase PVDF films are annealed at 70, 90, 110, 130 and $160^0$C for 5 hours to observe the changes in crystalline phases in PVDF films.

Figure 2 shows the XRD pattern of PVDF thin films, annealed at 70, 90, 110, 130 and $160^0$C for 5 hours. For comparison, the XRD pattern of non-annealed thin film is also mentioned in figure 2(a). From the XRD peak positions at different annealing conditions it is clear that γ-phase is mainly present in non-annealed thin films. At $160^0$C for 5 hours annealed condition, only α-phase is present in the film (figure 2(f)). In other annealing conditions β- or γ- phases are present along with α- phase. Peaks at 20.7 (200) and 20.8 (110) confirm the existence of maximum β-phase in PVDF thin films when (cast at room temperature from solution at $90^0$C) these are annealed at $90^0$C for five hours (figure 2(c)).

For further confirmation of phases in PVDF films are subjected to FTIR, Raman and DTA characterizations.

According to literature, the peaks at 411, 530, 615, 766, 795, 855, 974, 1149, 1210, 1383, 1402, 1432 and 1455$cm^{-1}$ are used for α-phase identification[21, 22]. The peaks at 511, 600, 840 and 1275$cm^{-1}$ are fingerprints of β-phase [21, 22]. The peak at 430, 778, 812, 832 and 1234$cm^{-1}$ show the presence of γ-phase without any combination with other phases [21, 22]. FTIR spectra of PVDF thin films in the range 400$cm^{-1}$ to 1750$cm^{-1}$ are shown in figure 4.

Prominent peaks at 435, 511, 600, 840, 1232, 1406, 1430 and 1455$cm^{-1}$ are observed in FTIR spectra of the non-annealed thin films (figure 3(a)). So from these peaks it is confirmed the presence of maximum percentage of γ- phase in the PVDF thin film (cast at room temperature from solution at $90^0$C). When thin films are annealed at $160^0$C for 5 hours at $2^0$C/min and cooled back to room temperature, then absorption peaks appear at 530, 615, 766, 854, 975, 1149, 1211, 1385, 1430 and 1455$cm^{-1}$ (figure 3(f)), indicating the complete conversion of thin film into α-phase. Figure 3(b)-(e) show the FTIR spectra of PVDF films, at annealing condition 70, 90, 110 and $130^0$C for 5 hours respectively. The maximum increase of β-phase is observed when thin film of PVDF is annealed at $90^0$C for 5hours (figure 3(c)). The increase in intensities of the peaks at 511, 600 and 1276$cm^{-1}$ indicates the increase of β- phase in PVDF thin films compared to other annealing conditions. In the region 400$cm^{-1}$ to 1000$cm^{-1}$ all atoms of PVDF are taking part in vibration. The vibrational modes, in the region 1000$cm^{-1}$ to 1330$cm^{-1}$ influence the dipole moment. The spontaneous polarization results from alignment of the dipole moment. So the vibrational modes in between 1000$cm^{-1}$ to 1330$cm^{-1}$ are responsible for the spontaneous polarization of β-phase PVDF and hence to ferroelectricity of PVDF and have large relative intensity in FTIR spectrum ( which is observed in figure 3 as per literature) [23].

The thin films were subjected for DTA analysis at a rate $5^0$C/min. Figure 4(a) shows four endothermic peaks at $128^0$C (broad), $168^0$C (diffused), $172^0$C and $192^0$C. From DSC and DTA studies it is clear that the melting temperature of α, β and γ phases of PVDF are 165-170$^0$C, 172-177$^0$C and 187-192$^0$C respectively [11, 19, 25, 26]. The small endothermic peak in figure 4(b) (compared to figure 4(a)) at 192$^0$C indicates the decrease of γ -phase percentage in the thin films at annealing condition 70$^0$C for 5 hours. The quantitative reduction of peak height at 128$^0$C (α-phase relaxation peak to β phase) [27] indicates the percentage of α-phase reduction at annealing condition 70$^0$C for 5 hours. The endothermic peak at 168$^0$C and 172$^0$C (both peaks are not well separated) indicates the presence of α- and β- phases in PVDF thin film. At annealing condition, 90$^0$C for 5 hours, the endothermic peaks at 128$^0$C and 192$^0$C subside to minimum (figure 4(c)). This points to increase of β-phase and substantial decrease of γ- phase and α- phase in the thin film. It has established the fact that the films annealed at 90$^0$C for 5 hours has maximum percentage of β-phase. The increase in the annealing temperature to 110$^0$C for same time period, results in increase of α-phase content in the thin film (figure 4(d)). The reappearance of endothermic α-relaxation peak at 128$^0$C and the prominent endothermic peak at 168$^0$C (melting temperature of α-phase) indicate the presence of α-phase in PVDF thin film, at annealing condition 110$^0$C for 5 hours. Thin films also contain β-phase, which is confirmed from diffused peak at 172$^0$C. At 160$^0$C for 5 hours (figure 4(f)), the PVDF thin films mainly contain α-phase. At annealing condition 160$^0$C for 5 hours, the large α relaxation peak at 128$^0$C and α-phase melting peak at 168$^0$C validate the presence of α-phase in PVDF thin film.

Raman spectroscopy provides more information about the conjugated structure and the chain skeleton of polymers. From figure 5, it is observed that there are gradual changes in the Raman spectra when the PVDF thin films are annealed at different temperatures for 5hours. The Raman bands at 796 and 811 in non-annealed PVDF films, correspond to α and γ phases respectively. The peak at 839cm$^{-1}$ shows the presence of both γ and β phases in PVDF film. The broad Raman peak at 882cm$^{-1}$ represents the combination of all three phases. For 110 and 130$^0$C annealing conditions an extra peak appears at 1131cm$^{-1}$ because conjugated C=C double bond formation in the PVDF thin films [24]. The PVDF thin films are completely converted to α-phase when these films are annealed at 160$^0$C for five hours (which is clear from the only Raman peak at 799cm$^{-1}$ appear in the spectrum).

The maximum percentage of β-phase PVDF has been obtained by annealing the PVDF thin film (cast using DMSO solution) at 90$^0$C for 5 hours. As per literature, for an applied field of 1200 kV/cm, the remnant polarization is 5μC/cm$^2$ at 0.001Hz. But at an applied field of 400kV/cm the β-PVDF shows polarization of 0.2μC/cm$^2$ at 0.001 Hz [28].

The figure 6 shows the hysteresis loops of unpoled β-phase PVDF for different frequencies at constant field 400kV/cm. When frequency of applied field 400kV/cm decreases from 50Hz to 1 Hz the ferroelectric polarization increases from 0.045μC/cm$^2$ to 0.186μC/cm$^2$. If the spontaneous polarization value at 400kV/cm is compared with the published literature values at 1Hz, the β-phase content in unpoled thin film (synthesized from DMSO solution and annealed at 90$^0$C for 5 hours) is above 95%.

**Why there is appreciable increase of β-phase content at annealing condition 90$^0$C for 5 hours?**

The temperature and time of crystallization determine the presence of predominant phases in PVDF. We have taken γ phase thin film with small percentage of α phase as initial film after casting. When thin film annealed at lower or near to α relaxation temperature (70-85$^0$C), the temperature is not sufficient high to destroy the crystalline order present in PVDF film. In this phase the viscocity is very high. So there is no change of phase when thin films annealed at 70$^0$C for 5 hours.

At annealing temperature 90$^0$C for 5 hours the viscocity of material decreases but still high enough to prevent the orientation of the crystals but the chain mobilty increases enough to reorganize the structure of conformers. Since 90$^0$C is just above α relaxation temperature the conversion rate from α to β phase occurs due to motion of conformers without considerable deformation of the crystals. At this temperature crystallization rate of β phase is also higher among other phases. So after annealing the PVDF film at 90$^0$C for 5 hours, the increase in β phase has been observed in it.

Above 90$^0$C and below 114$^0$C the viscocity of material decreases, the mobile fraction increases because of an oscillation motion of VDF groups of trans-zigzag chains with an amplitude of 10$^0$. So the films annealed at 110$^0$C for 5 hours consist of β phase with increase content of α phase.

As the temperature increases above 115$^0$C, the viscocity of PVDF decreases further to allow the active chain motion and chain reorientation in the crystalline region through the trans-gauche conformational exchange [29]. So due to deformation in crystalline region and reorientation of chains more stable α phase reappears in the PVDF films at high temperature annealing conditions. Above 160$^0$C PVDF completely converted to α phase.

## 4. Conclusion

In this report it has been observed that same solvent can yield all three phases of PVDF. By changing the solution temperature, different phases of PVDF can be obtained from same DMSO solvent. The annealing conditions (70, 90, 110, 130 and 160$^0$C for 5 hours) decide the presence of different phases and conversion of one phase to other phase of PVDF. The β-phase which is important for ferroelectric application has been obtained from γ-phase PVDF by suitable annealing. Maximum β- phase (above 95%) exist in the films when PVDF films are annealed at 90$^0$C for 5 hours. The γ phase completely converted to α-phase at 160$^0$C for 5 hours. The phase conversion at different annealing condition has been confirmed from XRD, FTIR, DTA, and Raman studies. Polarization measurement of unpoled thin film, having β phase content greater than 95%, shows a remnant polarization of 0.186μC/cm$^2$ at (400kV/cm) at an applied frequency of 1Hz.

We thank Sanjib Karmakar, Dr. V. G. Sathe, Pragya Pandit, Dr. Gurvinderjit Singh, Dr. S. M. Gupta and Indranil Bhoumik for their support in characterization.

**References:**

[1] Lando J. B., Olf H. G., Peterlin A., 1966 *J. Polym. Sci. Part A-1: polymer chemistry,* **4** 941.

[2] Hasegawa R., Takahashi Y., Chatani Y., Tadokoro H., 1972 *Polymer* **3** 600.

[3] Weinhold S., Litt M. H., Lando J. B., 1980 *Macromolucles* **13** 1178.

[4] Davis G. T., McKinney J. E., Broadhurst M. G., Roth S. C., 1978 *J. Appl. Phys.* **49** 4998.

[5] Dasgupta D. K., Doughty K., 1978 *J. Appl. Phys*. **49** 4601.

[6] Newman B. A., Yoon C. H., Pae K. D., 1979 *J. Appl. Phys.* 50 6095.

[7] Buonomenna M. G., Macchi P., Davoli M., Drioli E., 2007 *Euro. Polym. Jnl.* **43** 1557.

[8] He Xujiang, Yao Kui, 2006 *Appl. Phys. Lett.* **89** 112909.

[9] Mohammadi B., Yousefi A., Bellah S. M., 2007 *Polymer testing* **26** 42.

[10] Sajkiewicz P., Wasiak A., Goclowski Z., 1999 *Euro. Polym. Jnl.* **35,** 423.

[11] Salami A., Yousefi A. A., 2003 *Polymer Testing* **22** 699.

[12] Hasegawa R., Kobayashi M., Tadokoro H., 1972 *Polymer Journal* **3** 591.

[13] Okuda K., Yoshida T., Sugita M., Asahina M., 1967 *J. Polym. Sci. Part B: Polymer Letters* 5 465.

[14] Cortili G., Zerbi G., 1967 *Spectrochim Acta Part A: molecular spectroscopy* **23** 285.

[15] Gal'Perin Ye. L., Kosmynim B. P., Bychkov R. A., 1970 *Vysokomol Soe-din B* **12** 555.

[16] Park Youn Jung, Kang Yong Soo, Park Cheolmin, 2005 *Euro. Polym. Jnl*. **41** 1002.

[17] Welch G. J., Miller R. L., 1976 *J. Polym. Sci.: Polymer physics edition* **14** 1683.

[18] Prest Jr W. M., Luca D. J., 1975 *J. Appl. Phys*. **46** 4136.

[19] Prest Jr W. M., Luca D. J., 1978 *J. Appl. Phys*. **49** 5042.

[20] Y. Takahashi, A. Tadokoro, Macromolecules 13, 1318 (1980)

[21] Bormashenko Ye., Pogreb R., Stanevsky O., Bormashenko Ed., 2004 *Polymer Testing* **23** 791.

[22] Kobayashi M., Tashiro K., Tadokoro H., 1975 *Macromolecules* **8** 158.
11

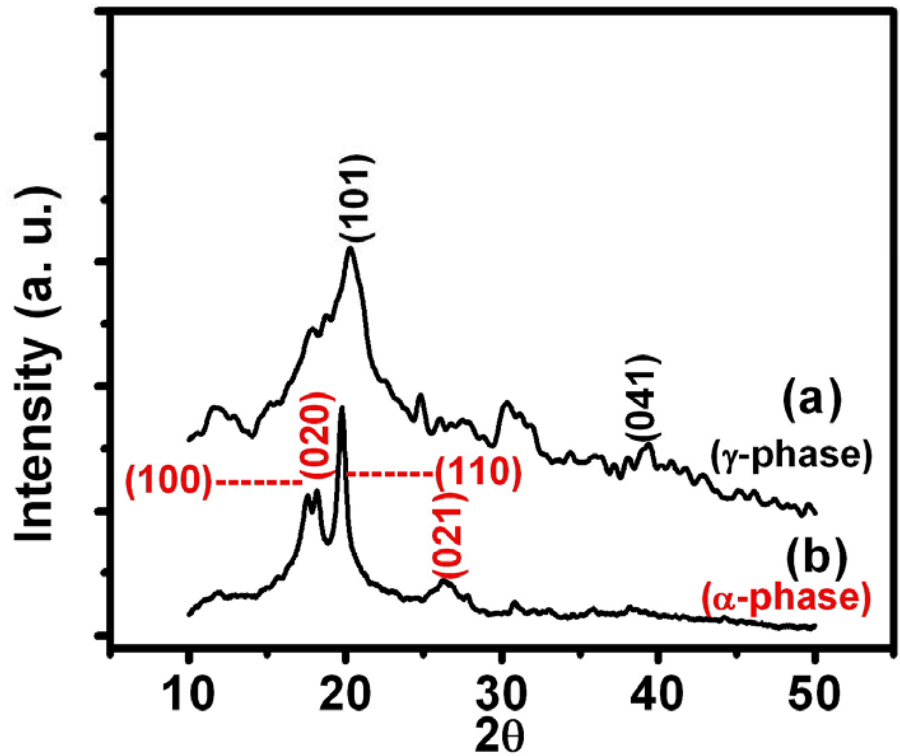

**Figure 1.** X-ray diffraction pattern of PVDF films (a) thin film cast at room temperature from 15wt% DMSO solution at $90^0$C. (b) thin film cast at room temperature from 15wt% DMSO solution at $50^0$C.

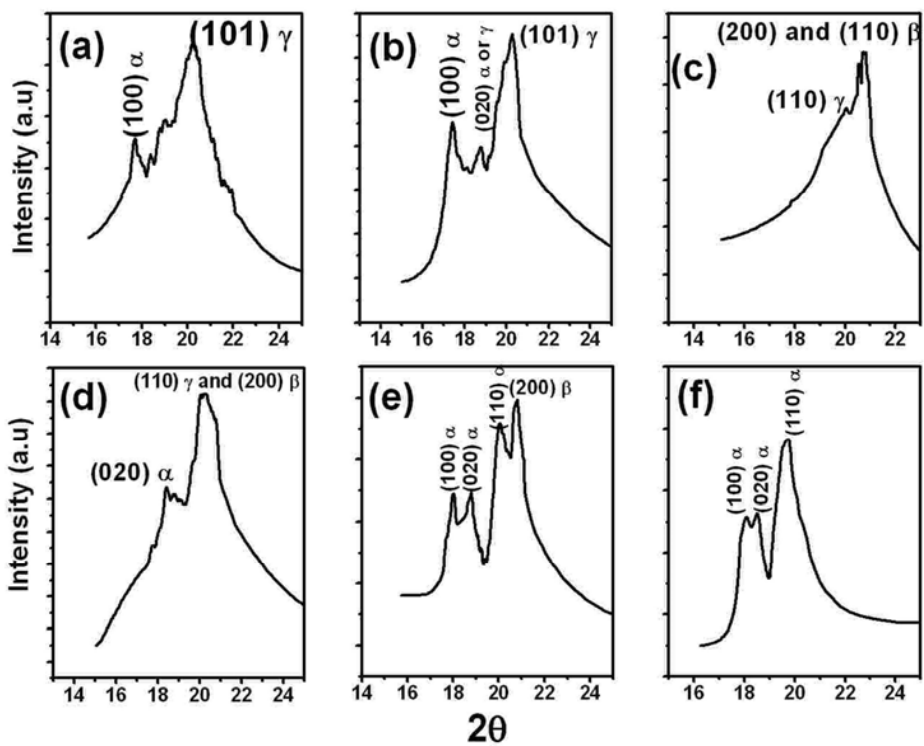

**Figure 2.** X-ray diffraction pattern of PVDF films (a) non-annealed, (b) annealed at 70$^0$C for 5 hours, (c) annealed at 90$^0$C for 5 hours, (d) annealed at 110$^0$C for 5 hours, (e) annealed at 130$^0$C for 5 hours, (f) annealed at 160$^0$C for 5 hours.

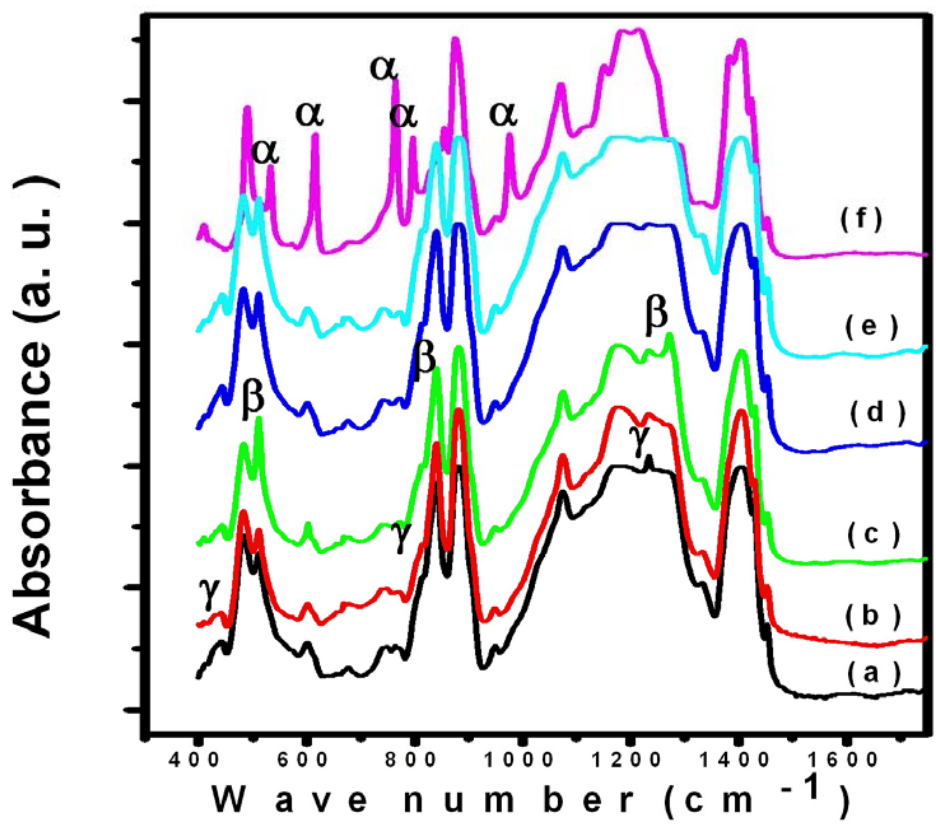

**Figure 3.** FTIR spectra of PVDF films: (a) non-annealed (b) annealed at $70^0$C for 5 hours, (c) annealed at $90^0$C for 5 hours, (d) annealed at $110^0$C for 5 hours, (e) annealed at $130^0$C for 5 hours, (f) annealed at $160^0$C for 5 hours.

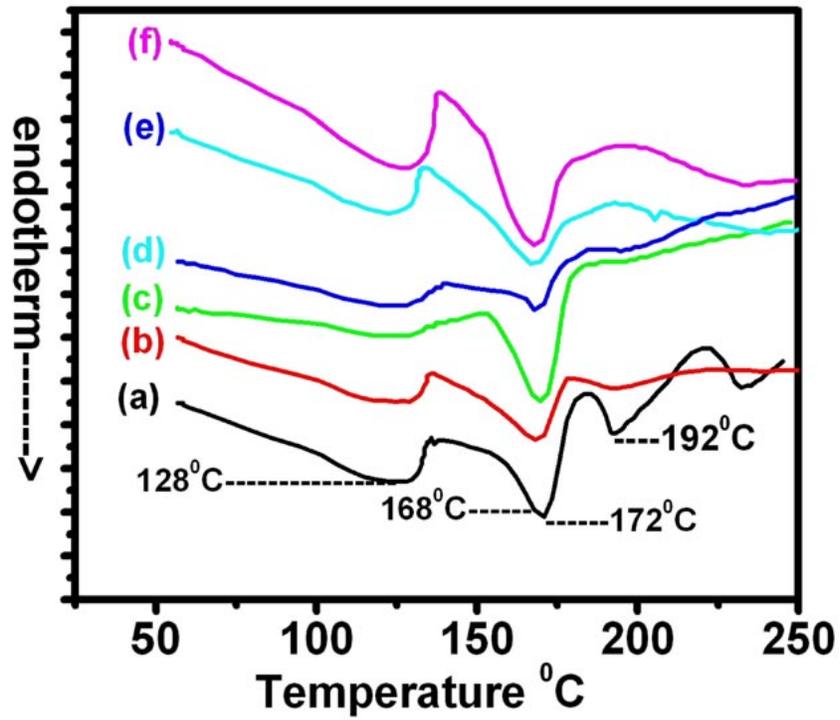

**Figure 4.** DTA thermo grams of PVDF films with heating rate 5$^0$C/min: (a) non-annealed (b) annealed at 70$^0$C for 5 hours, (c) annealed at 90$^0$C for 5 hours, (d) annealed at 110$^0$C for 5 hours, (e) annealed at 130$^0$C for 5 hours, (f) annealed at 160$^0$C for 5 hours.

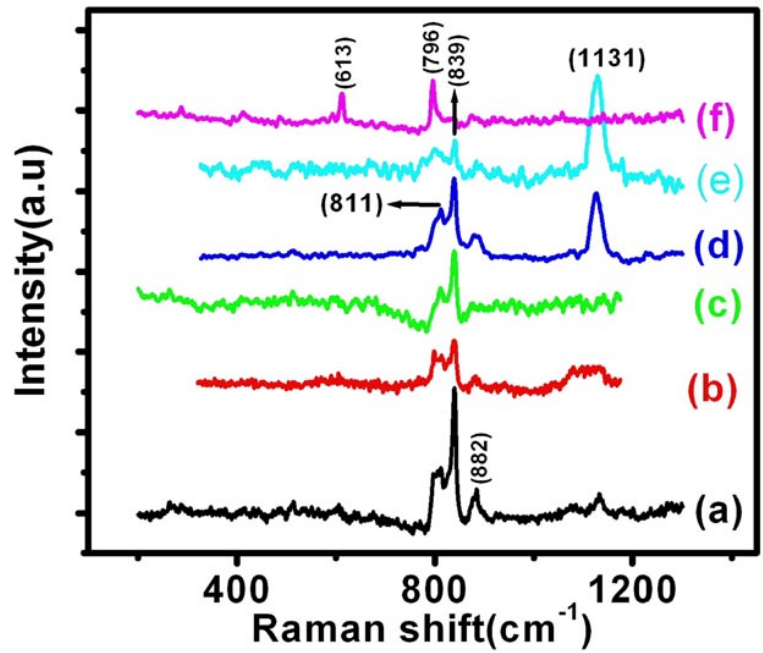

**Figure 5.** Raman spectra of PVDF films: (a) non-annealed (b) annealed at $70^0$C for 5 hours, (c) annealed at $90^0$C for 5 hours, (d) annealed at $110^0$C for 5 hours, (e) annealed at $130^0$C for 5 hours, (f) annealed at $160^0$C for 5 hours.

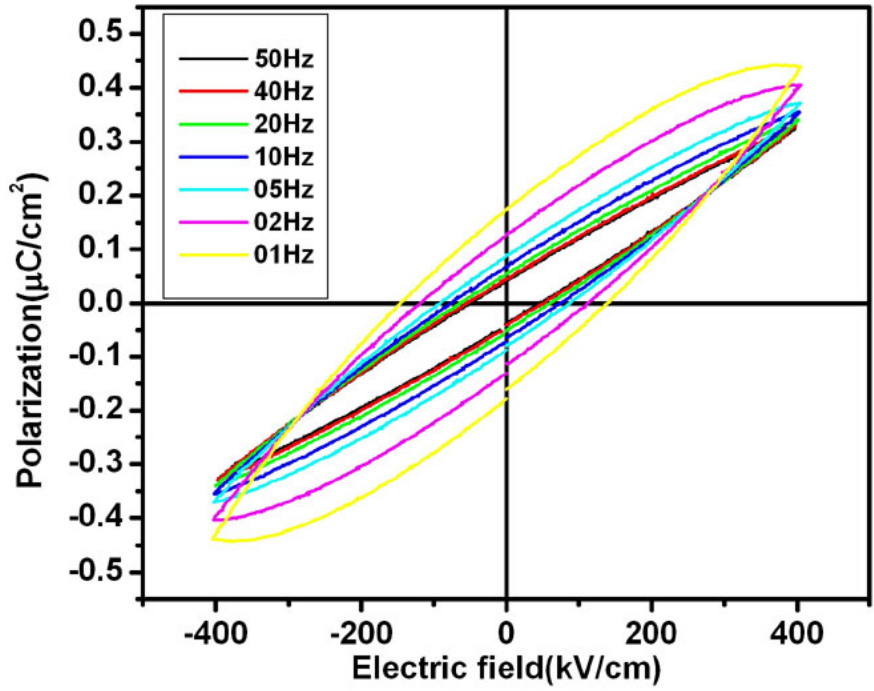

**Figure 6.** Hysteresis loop of unpoled transparent PVDF films (cast from DMSO solution and annealed at $90^0$C for 5 hours) at applied field 400kV/cm at different frequencies.